\documentclass[prd,showpacs,amsmath,amssymb]{revtex4}
\usepackage{graphicx}
\usepackage{dcolumn}
\usepackage{bm}

\begin{document}

\title{Path Integral of Bianchi I models in Loop Quantum Cosmology}
\author{Xiao Liu}
\email{lxiao@mail.bnu.edu.cn}
\affiliation{Department of Physics, Beijing Normal University, Beijing 100875, China}
\author{Fei Huang}
\affiliation{Department of Physics, Beijing Normal University, Beijing 100875, China}
\author{Jian-Yang Zhu}
\thanks{Author to whom correspondence should be addressed}
\email{zhujy@bnu.edu.cn}
\affiliation{Department of Physics, Beijing Normal University, Beijing 100875, China}
\date{\today}

\begin{abstract}
A path integral formulation of the Bianchi I models containing a massless scalar field in loop quantum cosmology is constructed. Following the strategy used in the homogenous and isotropic case, the calculation is extended to the simplest non-isotropic models according to the $\bar{\mu}$ and $\bar{\mu}^{\prime }$ scheme. It is proved from the path integral angle that the quantum dynamic lacks the full invariance with respect to fiducial cell scaling in the $\bar{\mu}$ scheme, but it does not in the $\bar{\mu}^{\prime }$ scheme. The investigation affirms the equivalence of the canonical approach and the path integral approach in loop quantum cosmology.
\end{abstract}

\maketitle

\section{Introduction}

Loop quantum gravity \cite{LQG} and the spinfoam formalism \cite{SF1,SF2} can be considered respectively as canonical quantization and covariant quantization of gravity just like what we did in ordinary quantum field theory. Applying the the techniques used in LQG to simple cosmology models, we get loop quantum cosmology \cite{LQC1}, which is a canonical version of quantum cosmology. Naturally, we desire to understand LQC conclusions from a path integral, i.e., covariant angle. Starting from the Hilbert space of LQC, there are two kinds of path integrals. One is integral over paths in configuration space \cite
{LQC-SF1,LQC-SF2,LQC-SF3,LQC-SF4}, which leads to a spinfoam like 'vertex summation' rather than a standard path integral format. The other is integral over phase space paths \cite{PATH1,PATH2,PATH3} from which we can get a typical continuous path integral that features continuous paths with weights given by the exponential of the phases.

In this paper, we will use phase space paths following the methods in the original work of A. Ashtekar, M. Campiglia and A. Henderson \cite{PATH1}, and
extend the LQC path integral formulation from homogenous and isotropic FRW model to the simplest homogenous Bianchi type I models \cite{BI1}%
. Up to now there are two kinds of 'loop regularization' of the gravitational part of the scalar constraint for the Bianchi I models in the literature, referred to as $\bar{\mu}$ and $\bar{\mu}^{\prime }$ scheme \cite {BI2,BI3}. Each of them has its own advantages and drawbacks. $\bar{\mu}$
scheme gives a difference equation in terms of affine variables and therefore the well-developed framework of the spatially flat-isotropic LQC
can be straightforwardly adopted. While the $\bar{\mu}^{\prime }$ scheme has better scaling properties \cite{BI3}, but the difference equation of this
scheme is very complex. Here we use both $\bar{\mu}$ and $\bar{\mu}^{\prime }$ schemes presented in \cite{BI2} to perform a path integral
formulation, and to explore their similarities and differences.

The organization of this paper is as follows. In the Sec. \ref{SEC2}, we present the quantum theory of Bianchi I modles in the volume and connection
representations. In Sec. \ref{SEC3}, we construct the path integral formalism of quantum Bianchi I models in phase space using $\bar{\mu}$ scheme, and the construction in $\bar{\mu}^{\prime }$ scheme will be presented in the Sec. \ref{SEC4}. Finally, we discuss the results obtained from those two
schemes in the Sec. \ref{SEC5}.

\section{LQC of the diagonal Bianchi I models}

\label{SEC2}

Bianchi cosmologies are homogeneous cosmological models, in which there is a foliation of spacetime $M=\Sigma \times \mathbb{R}$ such that $\Sigma $ is
space-like and there is a transitive isometry group freely acting on $\Sigma$. Thanks to these symmetry properties, in classical theory, after the
diagonalizing and rescaling processes of the Ashtekar variables (see \cite{BI1}), the only non-vanishing constraint(scalar constraint) for the Bianchi I
cosmology (coupled with a massless scalar field $\phi $) is given by
\begin{equation}
C=-\frac 1{8\pi G\gamma ^2}\left(c^1p_1c^2p_2+c^1p_1c^3p_3+c^2p_2c^3p_3\right) +\frac 12p_\phi ^2,
\label{constraint1}
\end{equation}
where $p_\phi $ is the momentum of the scalar field, while $c^i$ and $p_j$ are the reduced and rescaled Ashtekar variables. The Poisson brackets of $c^i$ and $p_j$ are
\begin{equation}
\{{c}^i,{p}_j\}=8\pi \gamma \,G\,\delta _j^i.
\end{equation}
In the isotropic models, there is a reflection symmetry $\Pi (p)=-p$, which corresponds to the
orientation reversal of triads. These are gouge transformations which leave the Hamiltonian constraint invariant. In Bianchi I case, we have three
reflections $\Pi _i$'s corresponding respectively to the reversal of one of the triad while leaving the other two fixed. The Hamiltonian flow is invariant under
the action of each $\Pi _i$ \cite{BI3}. Hence it suffices to restrict our attention to one of the octant of three $p_i$'s. In this paper, we focus on
the positive octant in which all three $p_i$'s are positive.

\subsection{$\bar{\mu}$ scheme}

\label{SEC2.1}

In the canonical quantization theory, i.e., loop quantum cosmology, there exists a comprehensive construction of the operator corresponding to the Eq. (\ref{constraint1}) in the full Loop quantum gravity scheme \cite{BI1}. The full LQG quantization proposes two kinds of quantum corrections to the scalar
constraint, one of which is the inverse triad correction by using a so-called Thiemann trick, while the other one is holonomy correction which means to correct the
curvature term by using a $SU(2)$ holonomy. In this paper, we start out setting the lapse function $N=|p_1p_2p_3|^{1/2}$ in the classical
theory as in \cite{BI3}. This treatment doesn't only make the Hamiltonian constraint simpler, but also avoids the use of inverse triad correction. Hence, we can
use the regularized constraint \cite{BI2} to replace Eq. (\ref{constraint1}), in which only the LQG effects from holonomy correction are taken into account:
\begin{equation}
C=-\frac 1{8\pi G\gamma ^2}\left[ \frac{\sin \left( {\bar{\mu}_1c^1}\right)}{\bar{\mu}_1}p_1\frac{\sin \left( {\bar{\mu}_2c^2}\right) }{\bar{\mu}_2}p_2+%
\frac{\sin \left( {\bar{\mu}_1c^1}\right) }{\bar{\mu}_1}p_1\frac{\sin \left({\bar{\mu}_3c^3}\right) }{\bar{\mu}_3}p_3+\frac{\sin \left( {\bar{\mu}_2c^2}%
\right) }{\bar{\mu}_2}p_2\frac{\sin \left( {\bar{\mu}_3c^3}\right) }{\bar{\mu}_3}p_3\right] +\frac{p_\phi ^2}2=0.  \label{constraint2}
\end{equation}

In $\bar{\mu}$ scheme, $\bar{\mu}_i=\sqrt{\Delta /|p_i|}$, where $\Delta =4\sqrt{3}\pi \gamma l_{Pl}^2$ is the 'area gap' \cite{AG}. Usually, we can
make an algebraic simplification by introducing new phase space variables:
\begin{equation}
\nu _i=\frac{{p_i}^{3/2}}{6\pi G},b_i=\frac{c_i}{\gamma \sqrt{p_i}},
\end{equation}
where the index $i$ does not sum over and their Poisson bracket is given by
\begin{equation}
\{b^i,\nu _j\}=2\delta _j^i.
\end{equation}
Then the kinematical Hilbert space of the gravitational part is given by
\begin{equation}
{\cal H}_{Kin}^{grav}=L^2({\mathbb{R}}_{Bohr},d\mu _{Bohr})^3,
\end{equation}
with the orthonormal basis $\left| \nu _1,\nu _2,\nu _3\right\rangle =\left|\nu _1\right\rangle \left| \nu _2\right\rangle \left| \nu _3\right\rangle$
satisfying
\begin{equation}
\left\langle \nu _1,\nu _2,\nu _3\right. \left| \tilde{\nu}_1,\tilde{\nu}_2,\tilde{\nu}_3\right\rangle =\delta _{\nu _1,\tilde{\nu}_1}\delta _{\nu _2,%
\tilde{\nu}_2}\delta _{\nu _3,\tilde{\nu}_3}.
\end{equation}
Any state $|\Psi (\nu _1,\nu _2,\nu _3)\rangle \in {\cal H}_{Kin}^{grav}$ can be decomposed in the orthonormal basis as
\begin{equation}
\left| \Psi (\nu _1,\nu _2,\nu _3)\right\rangle =\sum_{\nu _1,\nu _2,\nu _3}\Psi (\nu _1,\nu _2,\nu _3)\left| \nu _1,\nu _2,\nu _3\right\rangle .
\end{equation}

To define the constraint operator, we have to define two kinds of operators
first. One is the volume operator defined as
\begin{equation}
\hat{V}\left| \nu _1,\nu _2,\nu _3\right\rangle =6\pi G\gamma \sqrt{\Delta }|\nu _1\nu _2\nu _3|^{1/3}\left| \nu _1,\nu _2,\nu _3\right\rangle ,
\end{equation}
where $\hat{V}=|\hat{V}_1\hat{V}_2\hat{V}_3|^{1/3}$, and $\hat{V}_i\left| \nu_1,\nu _2,\nu _3\right\rangle =6\pi G\gamma \sqrt{\Delta }\nu _i\left| \nu
_1,\nu _2,\nu _3\right\rangle $, which is the operator corresponding to $p_i$. The other is the unitary shift operator
\begin{equation}
\hat{U}^i|\nu _i\rangle =\widehat{\exp \left( i\bar{\mu}_ic^i\right) \left|\nu _i\right\rangle }=\left| \nu _i+2\ell _0\hbar \right\rangle ,
\end{equation}
where $\ell _0$ is related to the 'area gap' by $\ell _0^2=\gamma ^2\Delta $, where $\gamma $ is the Barbero-Immirzi parameter. Replace the function $\sin
{\bar{\mu}_ic^i}$ in Eq. (\ref{constraint2}) by the operator $\widehat{\sin \bar{\mu}_ic^i}$ using the unitary shift operator defined above, we get
\begin{eqnarray}
\widehat{\sin (\bar{\mu}_ic^i)}\left| \nu _i\right\rangle &=&\frac 1{2i}\left( \widehat{\exp (i\bar{\mu}_ic^i)}-\widehat{\exp (-i\bar{\mu}_ic^i)}%
\right) \left| \nu _i\right\rangle  \nonumber \\
&=&\frac 1{2i}\left( \left| \nu _i+2\ell _0\hbar \right\rangle -\left| \nu_i-2\ell _0\hbar \right\rangle \right).
\end{eqnarray}
Then we fix the factor ordering ambiguity of Eq. (\ref{constraint2}), by defining the operator
\begin{align}
\hat{\Theta}_i|\nu _i\rangle :=& \frac 12\left( \frac{\widehat{\sin \bar{\mu}_ic^i}}{\sqrt{\Delta }}\hat{V}_i+\hat{V}_i\frac{\widehat{\sin \bar{\mu}_ic^i}%
}{\sqrt{\Delta }}\right) \left| \nu _i\right\rangle  \nonumber \\
=& -3i\pi G\gamma \left[ \left( \nu _i+\ell _0\hbar \right) \left| \nu_i+2\ell _0\hbar \right\rangle -\left( \nu _i-\ell _0\hbar \right) \left|
\nu _i-2\ell _0\hbar \right\rangle \right] .
\end{align}
Notice that $[\hat{\Theta}_i,\hat{\Theta}_j]|\nu _1,\nu _2,\nu _3\rangle =0$ for $i\neq j$.
The gravitational part of the scalar constraint operator can be written in terms of these operator as
\begin{equation}
\hat{C}^{grav}=-\frac 1{8\pi G\gamma ^2}\left( \hat{\Theta}_1\hat{\Theta}_2+\hat{\Theta}_2\hat{\Theta}_3+\hat{\Theta}_1\hat{\Theta}_3\right).
\end{equation}
The action of this operator on arbitrary basis element of the kinematical Hilbert space is \cite{BI4}
\begin{eqnarray}
&&\hat{C}^{grav}|\nu _1,\nu _2,\nu _3\rangle \nonumber \\
&=&\frac{9\pi G}8\left[ (\nu _1+1)(\nu _2+1)\left| \nu _1+2,\nu _2+2,\nu_3\right\rangle -(\nu _1-1)(\nu _2+1)\left| \nu _1-2,\nu _2+2,\nu
_3\right\rangle \right.  \nonumber \\
&&-(\nu _1+1)(\nu _2-1)\left| \nu _1+2,\nu _2-2,\nu _3\right\rangle +(\nu_1-1)(\nu _2-1)\left| \nu _1-2,\nu _2-2,\nu _3\right\rangle  \nonumber \\
&&+(\nu _2+1)(\nu _3+1)\left| \nu _1,\nu _2+2,\nu _3+2\right\rangle -(\nu_2-1)(\nu _3+1)\left| \nu _1,\nu _2-2,\nu _3+2\right\rangle  \nonumber \\
&&-(\nu _2+1)(\nu _3-1)\left| \nu _1,\nu _2+2,\nu _3-2\right\rangle +(\nu_2-1)(\nu _3-1)\left| \nu _1,\nu _2-2,\nu _3-2\right\rangle  \nonumber \\
&&+(\nu _1+1)(\nu _3+1)\left| \nu _1+2,\nu _2,\nu _3+2\right\rangle -(\nu_1-1)(\nu _3+1)\left| \nu _1-2,\nu _2,\nu _3+2\right\rangle  \nonumber \\
&&\left. -(\nu _1+1)(\nu _3-1)\left| \nu _1+2,\nu _2,\nu _3-2\right\rangle+(\nu _1-1)(\nu _3-1)\left| \nu _1-2,\nu _2,\nu _3-2\right\rangle \right] ,
\label{constraint3}
\end{eqnarray}
where the $\ell _0\hbar $ terms are omitted for short. From this equation we can see that in the $\bar{\mu}$ scheme, the space of solutions to the
quantum constraint is very similar to the isotropic case. $\nu _1,\nu _2,\nu _3$ are discrete variables and they are supported on a specific
superselections respectively. They can take the values of $\nu _i=(\varepsilon _i+2n_i)\ell _0\hbar $, where the parameter $\varepsilon_i\in [0,1]$, and $n_i\in \mathbb{N}$. As mentioned previously, we focus on the positive octant here. This precondition simplify the calculation but do not
affect the physical meaning. If we consider the general situation, a factor $sgn(p_i)$ will appear in the expression of $\hat{\Theta}_i$. And as a function on phase space, it does not commute with $\sin\bar{\mu}_ic^i$, hence their product as operators is not symmetric. It is necessary
to realize this point, although our simplification makes its role less important. For the details about this situation, see \cite{BI6}.

\subsection{$\bar{\mu}^{\prime}$ scheme}

\label{SEC2.2} We still use the regularized constraint Eq. (\ref{constraint2}) in $\bar{\mu}^{\prime }$ scheme. Only $\bar{\mu}_i$ changes to $\bar{\mu}^{\prime }_i$ in this situation, and $\bar{\mu}_i^{\prime }$ takes the form
\begin{equation}
\bar{\mu}_1^{\prime }=\sqrt{\frac{|p_1|\Delta }{|p_2p_3|}},\quad \bar{\mu}_2^{\prime }=\sqrt{\frac{|p_2|\Delta }{|p_1p_3|}},\quad \bar{\mu}_3^{\prime
}=\sqrt{\frac{|p_3|\Delta }{|p_1p_2|}}.  \label{mubarp}
\end{equation}
Owning to this condition, it's inconvenient to still use $\nu _i$ as configuration variables. By introducing new variables $\lambda _i$,
\begin{equation}
\lambda _i=\frac{\sqrt{p_i}}{(4\pi G)^{1/3}},
\end{equation}
the algebra of the $\bar{\mu}^{\prime }$ scheme could be much more simplified \cite{BI3}. The variable conjugate to $\lambda _i$ is
\begin{equation}
k^i=\frac{2\sqrt{p_i}c^i}{\gamma (4\pi G)^{2/3}}.  \label{k}
\end{equation}
We can prove that the resulting Poisson bracket between $k^i$ and $\lambda _j$ is
\begin{equation}
\{k^i,\lambda _j\}=2\delta _j^i.
\end{equation}
Then the new orthonormal basis is $|\lambda _1,\lambda _2,\lambda _3\rangle$ in ${\cal H}_{Kin}^{grav}$, and the fundamental volume operator and unitary
shift are given by
\begin{equation}
\hat{V}\left| \lambda _1,\lambda _2,\lambda _3\right\rangle =2\pi G\gamma\sqrt{\Delta }v\left| \lambda _1,\lambda _2,\lambda _3\right\rangle ,
\label{VE}
\end{equation}
and
\begin{equation}
\hat{E_i}|\lambda _i\rangle =\widehat{\exp \left( i\bar{\mu}_ic^i\right)\left| \lambda _i\right\rangle }=\left| \lambda _i+\frac{\ell _0\hbar }{
\lambda _j\lambda _k}\right\rangle.
\end{equation}
Here, $v=2\lambda _1\lambda _2\lambda _3$ is the configuration variable related to the volume of the elementary cell, and the factor $2$ ensures
that this $v$ reduces to the $v$ used in the isotropic case. We can also use $\lambda _i,\lambda _j$, and $v$ as the basic configuration variables to substitute for
$\lambda _1,\lambda _2$, and $\lambda _3$. The indices $i,j,k=1,2,3$ appeared in Eq. ({\ref{VE}}) are required to be different with each other, and, as usual, the index
$i$ does not sum over. Having these operators in hands, we could get the gravitational part of the scalar constraint operator. Its action on
arbitrary basis element of the new kinematical Hilbert space is \cite{BI3}
\begin{eqnarray}
\hat{C}^{grav}\left| \lambda _1,\lambda _2,v\right\rangle  &=&\frac{\pi G}2\sqrt{v}\left[ (v+2)\sqrt{v+4}\left| \lambda _1,\lambda _2,v\right\rangle
_4^{+}-(v+2)\sqrt{v}\left| \lambda _1,\lambda _2,v\right\rangle_0^{+}\right.   \nonumber \\
&&\left. -(v-2)\sqrt{v}|\lambda _1,\lambda _2,v\rangle _0^{-}+(v-2)\sqrt{|v-4|}|\lambda _1,\lambda _2,v\rangle _4^{-}\right] ,
\label{mubar'C}
\end{eqnarray}
where
\begin{eqnarray}
\left| \lambda _1,\lambda _2,v\right\rangle _4^{\pm } &=&\left| \frac{v\pm 4}{v\pm 2}\cdot \lambda _1,\frac{v\pm 2}v\cdot \lambda _2,v\pm 4\right\rangle
+\left| \frac{v\pm 4}{v\pm 2}\cdot \lambda _1,\lambda _2,v\pm 4\right\rangle \nonumber \\
&&+\left| \frac{v\pm 2}v\cdot \lambda _1,\frac{v\pm 4}{v\pm 2}\cdot \lambda_2,v\pm 4\right\rangle +\left| \frac{v\pm 2}v\cdot \lambda _1,\lambda
_2,v\pm 4\right\rangle   \nonumber \\
&&+\left| \lambda _1,\frac{v\pm 2}v\cdot \lambda _2,v\pm 4\right\rangle +\left| \lambda _1,\frac{v\pm 4}{v\pm 2}\cdot \lambda _2,v\pm
4\right\rangle , \label{mubar'C1}
\end{eqnarray}
and
\begin{eqnarray}
\left| \lambda _1,\lambda _2,v\right\rangle _0^{\pm } &=&\left| \frac{v\pm 2}v\cdot \lambda _1,\frac v{v\pm 2}\cdot \lambda _2,v\right\rangle +\left|
\frac{v\pm 2}v\cdot \lambda _1,\lambda _2,v\right\rangle   \nonumber \\
&&+\left| \frac v{v\pm 2}\cdot \lambda _1,\frac{v\pm 2}v\cdot \lambda_2,v\right\rangle +\left| \frac v{v\pm 2}\cdot \lambda _1,\lambda
_2,v\right\rangle   \nonumber \\
&&+\left| \lambda _1,\frac v{v\pm 2}\cdot \lambda _2,v\right\rangle +\left|\lambda _1,\frac{v\pm 2}v\cdot \lambda _2,v\right\rangle .
\label{mubar'C2}
\end{eqnarray}
As what we did in Eq. (\ref{constraint3}), the $\ell _0\hbar $ terms are omitted here. From Eq. (\ref{mubar'C1}) and Eq. (\ref{mubar'C2}), we can see that
the variable $v$ is fine, and the wave function only involves $\left( v-4\ell_0\hbar \right) ,v,\left( v+4\ell _0\hbar \right) $ terms. It is exactly
the same as in the isotropic case. On the contrary, the situation for $\lambda _1,\lambda _2$ is much different. We see that they depend only on the
value of $v$. This dependence is through fractional factors whose denominator is two or four units bigger or smaller than the numerator. This lead to
the consequence that the iterative action of the constraint operator derives only to states whose quantum numbers $\lambda_i$ are of the form 
$\lambda_i=\omega_{\epsilon}\lambda^{\prime}_i$, here $\lambda^{\prime}_i$ is the initial value of $\lambda_i$ and $\epsilon$ is a constant number 
that $v=\epsilon+4n\ell_0\hbar, n\in\mathbb{N}$. $\omega_{\epsilon}$ belonging to the set \cite{BI5}
\begin{equation}
\Omega_{\epsilon}=\left\{\left(\frac{\epsilon-2}{\epsilon}\right)^z\prod_{m,n\in\mathbb{N}}\left(\frac{\epsilon+2m}{\epsilon+2n}\right)^{k^m_n}\right\},
\end{equation}
where $k^m_n\in\mathbb{N}$, and $z\in\mathbb{Z}$ if $\epsilon>2$, while $z=0$ when $\epsilon\leq2$. The discrete set $\Omega_{\epsilon}$ is countably infinite and turns out to be dense in the positive real line. The proof of this statement see Appendix D3 of \cite{BI5}. So, On the one hand, $\lambda_i$ is superselected in separable sectors, have countable numbers of values. On the other hand, the dependence on the fractional factors makes it can take values in entire positive real line.

\section{Phase space path integral in $\bar{\mu}$ scheme}

\label{SEC3}

In this section, we calculate the extraction amplitude $A(\vec{\nu}_f,\phi_f;\vec{\nu}_i,\phi _i)$ for Bianchi I cosmology (here $\vec{\nu}$ is short
for $\{\nu _1,\nu _2,\nu _3\}$). As we know, in the 'timeless' framework of LQC, the whole information of the quantum dynamics is encoded in $A(\vec{\nu}%
_f,\phi _f;\vec{\nu}_i,\phi _i)$, which is the transition amplitude for our phase space path integral just as in ordinary quantum mechanics. We can express it as \cite{LQC-SF1}
\begin{equation}
A(\vec{\nu}_f,\phi _f;\vec{\nu}_i,\phi _i)=\int d\alpha \left\langle \vec{\nu%
}_f,\phi _f\right. \left| e^{\frac i\hbar \alpha \hat{C}}\right| \left. \vec{%
\nu}_i,\phi _i\right\rangle ,
\end{equation}
where the total constraint operator $\hat{C}$ is composed of two parts: the
gravitational part and the scalar field part
\begin{equation}
\hat{C}=\hat{C}^{grav}+\hat{C}^{matt}.  \label{totalC}
\end{equation}
The two operators $\hat{C}^{grav}$ and $\hat{C}^{matt}$ are commutative, so they can act on their own Hilbert space, ${\cal H}_{Kin}^{grav}$ and
${\cal H}_{Kin}^{matt}$, respectively. Then following the spirit of \cite{PATH1}, decompose the fictitious evolution into $N$ evolutions of length
$\epsilon =1/N$, and insert complete basis in between each factor, we get
\begin{equation}
\left\langle \vec{\nu}_f,\phi _f\right. \left| e^{\frac i\hbar \alpha \hat{C}%
}\right| \left. \vec{\nu}_i,\phi _i\right\rangle =\sum_{\vec{\nu}%
_{N-1}\cdots \vec{\nu}_1}\int d\phi _{N-1}\cdots d\phi _1\left\langle \vec{%
\nu}_N,\phi _N\right. \left| e^{\frac i\hbar \epsilon \alpha \hat{C}}\right|
\left. \vec{\nu}_{N-1},\phi _{N-1}\right\rangle \cdots \left\langle \vec{\nu}%
_1,\phi _1\right. \left| e^{\frac i\hbar \epsilon \alpha \hat{C}}\right|
\left. \vec{\nu}_0,\phi _0\right\rangle ,
 \label{amp1}
\end{equation}
where $\left\langle \vec{\nu}_N,\phi _N\right| \equiv \left\langle \vec{\nu}%
_f,\phi _f\right| $ and $\left| \vec{\nu}_0,\phi _0\right\rangle \equiv
\left| \vec{\nu}_i,\phi _i\right\rangle $.

Firstly, we need to calculate the $n$-th term appearing in above expression. Using the Eq. (\ref{totalC}), it can be stated as
\begin{eqnarray}
&&\left\langle \nu _{n+1}^1,\nu _{n+1}^2,\nu _{n+1}^3,\phi _{n+1}\right.
\left| e^{\frac i\hbar \epsilon \alpha \hat{C}}\right| \left. \nu _n^1,\nu
_n^2,\nu _n^3,\phi _n\right\rangle   \nonumber \\
&=&\left\langle \phi _{n+1}\right. \left| e^{\frac i\hbar \varepsilon \alpha
\hat{C}^{matt}}\right| \left. \phi _n\right\rangle \left\langle \nu
_{n+1}^1,\nu _{n+1}^2,\nu _{n+1}^3\right. \left| e^{\frac i\hbar \epsilon
\alpha \hat{C}^{grav}}\right| \left. \nu _n^1,\nu _n^2,\nu _n^3\right\rangle
.
\end{eqnarray}
As usual, the matter Hilbert space is the standard one, ${\cal H}_{Kin}^{matt}=L^2(\mathbb{R},d\phi )$. By an ordinary quantum mechanics like
calculation, the scalar field factor is
\begin{equation}
\left\langle \phi _{n+1}\right. \left| e^{\frac i\hbar \varepsilon \alpha
\hat{C}^{matt}}\right| \left. \phi _n\right\rangle =\left\langle \phi
_{n+1}\right. \left| e^{\frac i\hbar \varepsilon \alpha \hat{p}^2}\right|
\left. \phi _n\right\rangle =\int \frac{dp_n}{2\pi }e^{\frac i\hbar
p_n\left( \phi _{n+1}-\phi _n\right) +\frac i\hbar \epsilon \alpha p_n^2},
\label{Cmatt}
\end{equation}
and the gravitational factor is
\begin{eqnarray}
&&\left\langle \nu _{n+1}^1,\nu _{n+1}^2,\nu _{n+1}^3\right. \left| e^{\frac %
i\hbar \epsilon \alpha \hat{C}^{grav}}\right| \left. \nu _n^1,\nu _n^2,\nu
_n^3\right\rangle   \nonumber \\
&=&\delta _{\nu _{n+1}^1,\nu _n^1}\delta _{\nu _{n+1}^2,\nu _n^2}\delta
_{\nu _{n+1}^3,\nu _n^3}+\frac i\hbar \epsilon \alpha \left\langle \nu
_{n+1}^1,\nu _{n+1}^2,\nu _{n+1}^3\right. \left| \hat{C}^{grav}\right|
\left. \nu _n^1,\nu _n^2,\nu _n^3\right\rangle +{\cal O}(\epsilon ^2),
\end{eqnarray}
where we made an expansion in $\epsilon $, and ${\cal O}(\epsilon ^2)$ is the high order terms of the expansion. Using Eq. (\ref{constraint2}), the matrix element of $\hat{C}^{grav}$ is
\begin{eqnarray}
&&\left\langle \nu _{n+1}^1,\nu _{n+1}^2,\nu _{n+1}^3\right. \left| \hat{C}%
^{grav}\right| \left. \nu _n^1,\nu _n^2,\nu _n^3\right\rangle   \nonumber
\label{expC} \\
&=&\frac{9\pi G}8\left[ \frac{\nu _{n+1}^1+\nu _n^1}2\frac{\nu _{n+1}^2+\nu
_n^2}2(\delta _{\nu _{n+1}^1,\nu _n^1+2}\delta _{\nu _{n+1}^2,\nu
_n^2+2}\delta _{\nu _{n+1}^3,\nu _n^3}-\delta _{\nu _{n+1}^1,\nu
_n^1-2}\delta _{\nu _{n+1}^2,\nu _n^2+2}\delta _{\nu _{n+1}^3,\nu
_n^3}\right.   \nonumber \\
&&\left. -\delta _{\nu _{n+1}^1,\nu _n^1+2}\delta _{\nu _{n+1}^2,\nu
_n^2-2}\delta _{\nu _{n+1}^3,\nu _n^3}+\delta _{\nu _{n+1}^1,\nu
_n^1-2}\delta _{\nu _{n+1}^2,\nu _n^2-2}\delta _{\nu _{n+1}^3,\nu
_n^3}\right)   \nonumber \\
&&+\frac{\nu _{n+1}^2+\nu _n^2}2\frac{\nu _{n+1}^3+\nu _n^3}2\left( \delta
_{\nu _{n+1}^2,\nu _n^2+2}\delta _{\nu _{n+1}^3,\nu _n^3+2}\delta _{\nu
_{n+1}^1,\nu _n^1}-\delta _{\nu _{n+1}^2,\nu _n^2-2}\delta _{\nu
_{n+1}^3,\nu _n^3+2}\delta _{\nu _{n+1}^1,\nu _n^1}\right.   \nonumber \\
&&\left. -\delta _{\nu _{n+1}^2,\nu _n^2+2}\delta _{\nu _{n+1}^3,\nu
_n^3-2}\delta _{\nu _{n+1}^1,\nu _n^1}+\delta _{\nu _{n+1}^2,\nu
_n^2-2}\delta _{\nu _{n+1}^3,\nu _n^3-2}\delta _{\nu _{n+1}^1,\nu
_n^1}\right)   \nonumber \\
&&+\frac{\nu _{n+1}^1+\nu _n^1}2\frac{\nu _{n+1}^3+\nu _n^3}2\left( \delta
_{\nu _{n+1}^1,\nu _n^1+2}\delta _{\nu _{n+1}^3,\nu _n^3+2}\delta _{\nu
_{n+1}^2,\nu _n^2}-\delta _{\nu _{n+1}^1,\nu _n^1-2}\delta _{\nu
_{n+1}^3,\nu _n^3+2}\delta _{\nu _{n+1}^2,\nu _n^2}\right.   \nonumber \\
&&\left. \left. -\delta _{\nu _{n+1}^1,\nu _n^1+2}\delta _{\nu _{n+1}^3,\nu
_n^3-2}\delta _{\nu _{n+1}^2,\nu _n^2}+\delta _{\nu _{n+1}^1,\nu
_n^1-2}\delta _{\nu _{n+1}^3,\nu _n^3-2}\delta _{\nu _{n+1}^2,\nu
_n^2}\right) \right].
\end{eqnarray}
As demonstrated before, we choose a specific superselection of $\nu _i$ such that $\nu _i=2n_i\ell _0\hbar $, where $n_i\in \mathbb{N}$. Then we can use the
Fourier expansion formula of the Kronecker delta
\begin{equation}
\delta {\nu ^{\prime }\nu }=\frac{\ell _0}\pi \int_0^{\pi /\ell
_0}dbe^{-ib\left( \nu ^{\prime }-\nu \right) /2\hbar },
\label{Kroneckerdelta}
\end{equation}
where, obviously, $b$ is the conjugate variable to $\nu $ and take values in the
range $(0,\pi /\ell _0)$. By using Eq. (\ref{Kroneckerdelta}), then Eq. (\ref{expC})
can be expressed as
\begin{eqnarray}
&&\left\langle \nu _{n+1}^1,\nu _{n+1}^2,\nu _{n+1}^3\right. \left| e^{\frac %
i\hbar \epsilon \alpha \hat{C}^{grav}}\right| \left. \nu _n^1,\nu _n^2,\nu
_n^3\right\rangle   \nonumber \\
&=&\left( \frac{\ell _0}\pi \right) ^3\int d\vec{b}_{n+1}e^{-\frac i\hbar
\frac{b_{n+1}^1(\nu _{n+1}^1-\nu _n^1)}2}\cdot e^{-\frac i\hbar \frac{%
b_{n+1}^2(\nu _{n+1}^2-\nu _n^2)}2}\cdot e^{-\frac i\hbar \frac{%
b_{n+1}^3(\nu _{n+1}^3-\nu _n^3)}2}  \nonumber \\
&&\times \left\{ 1-\frac i\hbar \epsilon \alpha \frac 92\pi G\left[ \frac{%
\nu _{n+1}^1+\nu _n^1}2\frac{\nu _{n+1}^2+\nu _n^2}2\sin \ell
_0b_{n+1}^1\sin \ell _0b_{n+1}^2\right. \right.   \nonumber \\
&&\left. \left. +\frac{\nu _{n+1}^2+\nu _n^2}2\frac{\nu _{n+1}^3+\nu _n^3}2%
\sin \ell _0b_{n+1}^2\sin \ell _0b_{n+1}^3+\frac{\nu _{n+1}^1+\nu _n^1}2%
\frac{\nu _{n+1}^3+\nu _n^3}2\sin \ell _0b_{n+1}^1\sin \ell _0b_{n+1}^3+%
{\cal O}(\epsilon ^2)\right] \right\} .
\end{eqnarray}
Now according to the calculation before, we could put the scalar field part
and gravitational part together, and Eq. (\ref{amp1}) takes the form
\begin{eqnarray}
&&\left\langle \vec{\nu}_f,\phi _f\right. \left| e^{i\alpha \hat{C}}\right|
\left. \vec{\nu}_i,\phi _i\right\rangle   \nonumber \\
&=&\sum_{\nu _{N-1}\cdots \nu _1}\left( \frac{\ell _0}\pi \right) ^{3N}\int d%
\vec{b}_N\cdots d\vec{b}_1\cdot \left( \frac 1{2\pi }\right) ^{3N}\int
dp_N\cdots dp_1e^{\frac i\hbar S_N}+{\cal O}(\epsilon ^2),
\end{eqnarray}
where
\begin{eqnarray}
S_N &=&\epsilon \sum_{n=0}^{N-1}\left( p_{n+1}\frac{\phi _{n+1}-\phi _n}%
\epsilon -\frac{b_{n+1}^1}2\frac{\nu _{n+1}^1-\nu _n^1}\epsilon -\frac{%
b_{n+1}^2}2\frac{\nu _{n+1}^2-\nu _n^2}\epsilon -\frac{b_{n+1}^3}2\frac{\nu
_{n+1}^3-\nu _n^3}\epsilon \right)   \nonumber \\
&&+\alpha \left[ p_n^2-\frac 92\pi G\left( \frac{\nu _{n+1}^1+\nu _n^1}2%
\frac{\nu _{n+1}^2+\nu _n^2}2\sin \ell _0b_{n+1}^1\sin \ell
_0b_{n+1}^2\right. \right.   \nonumber \\
&&\left. \left. + \frac{\nu _{n+1}^2+\nu _n^2}2\frac{\nu _{n+1}^3+\nu
_n^3}2\sin \ell _0b_{n+1}^2\sin \ell _0b_{n+1}^3 + \frac{\nu
_{n+1}^1+\nu _n^1}2\frac{\nu _{n+1}^3+\nu _n^3}2\sin \ell _0b_{n+1}^1\sin
\ell _0b_{n+1}^3\right) \right].
\end{eqnarray}
This is the 'discrete-time action' of the Bianchi I cosmology. Just like in \cite{PATH1}, the final step is to take the limit $N\rightarrow \infty $, and
because the variable $\nu _i$ is discrete, it is impossible to interpret the $(\nu _{n+1}^i-\nu _n^i)/\epsilon $ as a derivative, we should transform the
terms $(\nu _{n+1}^i-\nu _n^i)/\epsilon $ into $(b_{n+1}^i-b_n^i)/\epsilon $:
\begin{equation}
\frac 12\left( \nu _Nb_N-\nu _0b_1\right) =\frac 12\sum_{n=0}^{N-1}\left(
b_{n+1}\nu _{n+1}-\nu _nb_n\right).
\end{equation}
Then
\begin{equation}
\epsilon \sum_{n=0}^{N-1}\left[ -\frac{b_{n+1}}2\frac{\nu _{n+1}-\nu _n}%
\epsilon \right] =\epsilon \sum_{n=0}^{N-1}\left[ \frac{\nu _n}2\frac{%
b_{n+1}-b_n}\epsilon \right] +\frac 12(b_1\nu _0-b_N\nu _N).
\end{equation}
Using this transformation, $S_N$ can be rewritten as
\begin{eqnarray}
S_N &=&\epsilon \left\{ p_{n+1}\frac{\phi _{n+1}-\phi _n}\epsilon +\frac{\nu
_n^1}2\frac{b_{n+1}^1-b_n^1}\epsilon +\frac{\nu _n^2}2\frac{b_{n+1}^2-b_n^2}%
\epsilon +\frac{\nu _n^3}2\frac{b_{n+1}^3-b_n^3}\epsilon \right.   \nonumber
\\
&&+\alpha \left[ p_n^2-\frac 92\pi G\left( \frac{\nu _{n+1}^1+\nu _n^1}2%
\frac{\nu _{n+1}^2+\nu _n^2}2\sin \ell _0b_{n+1}^1\sin \ell
_0b_{n+1}^2\right. \right.   \nonumber \\
&&\left. \left. \left. + \frac{\nu _{n+1}^2+\nu _n^2}2\frac{\nu
_{n+1}^3+\nu _n^3}2\sin \ell _0b_{n+1}^2\sin \ell _0b_{n+1}^3+ \frac{%
\nu _{n+1}^1+\nu _n^1}2\frac{\nu _{n+1}^3+\nu _n^3}2\sin \ell
_0b_{n+1}^1\sin \ell _0b_{n+1}^3\right) \right] \right\}   \nonumber \\
&&-\frac 12(\vec{\nu}_f\vec{b}_f-\vec{\nu}_i\vec{b}_i).
\end{eqnarray}
Finally, take the limit $N\rightarrow \infty $, we have
\begin{equation}
A(\vec{\nu}_f,\phi _f;\vec{\nu}_i,\phi _i)=\int \alpha \int [D\nu _q(\tau
)][Db_q(\tau )][Dp(\tau )][D\phi (\tau )]e^{\frac i\hbar \overline{S}},
\end{equation}
where
\begin{eqnarray}
\overline{S} &=&\int_0^1d\tau \left\{ p\dot{\phi}+\frac 12\vec{\nu}\cdot
\dot{\vec{b}}-\alpha \left[ p^2-\frac 92\pi G(\nu ^1\nu ^2\sin \ell _0b^1\sin \ell
_0b^2+\nu ^2\nu ^3\sin \ell _0b^2\sin \ell _0b^3\right. \right.   \nonumber
\label{action1} \\
&&\left. \left. +\nu ^1\nu ^3\sin \ell _0b^1\sin \ell _0b^3)\right] -\frac 12%
(\vec{\nu}_f\cdot \vec{b}_f-\vec{\nu}_i\cdot \vec{b}_i)\right\} .
\end{eqnarray}
As in the isotropic case, we use the subscript $q$ here to emphasize that now we can take a sum in the geometrical
sector including only the 'quantum paths'. Also, we use the same trick to transform it to the familiar format
\begin{equation}
\frac \pi {\ell _0}\sum_{\nu _n}\int_0^{\pi /\ell _0}db_n\rightarrow
\int_{-\infty }^\infty d\nu _n\int_{-\infty }^\infty db_n.  \label{trick}
\end{equation}
Using Eq. (\ref{trick}), we get the final expression of the extraction
amplitude
\begin{equation}
A(\vec{\nu}_f,\phi _f;\vec{\nu}_i,\phi _i)=\int d\alpha \int [D\nu (\tau
)][Db(\tau )][Dp(\tau )][D\phi (\tau )]e^{\frac i\hbar S},
\end{equation}
where
\begin{eqnarray}
S &=&\int_0^1d\tau \left\{ p\dot{\phi}-\frac 12\vec{b}\cdot \dot{\vec{\nu}}-\alpha
\left[ p^2\right. \right.   \nonumber  \label{muS} \\
&&\left. -\frac 92\pi G(\nu ^1\nu ^2\sin \ell _0b^1\sin \ell _0b^2+\nu ^2\nu
^3\sin \ell _0b^2\sin \ell _0b^3\right.   \nonumber \\
&&\left. \left. +\nu ^1\nu ^3\sin \ell _0b^1\sin \ell _0b^3)\right] \right\}
.
\end{eqnarray}
This is the path integral formulation and its action we desired. Because we used Eq. (\ref{trick}), the integration about $\nu _i$ and $b_i$ are taken
from $-\infty $ to $\infty $, with the boundary terms in Eq. (\ref{action1}) canceling each other out. And then we are able to integrate over all paths in the classical
phase space as in usual path integrals.

\section{Phase space path integral in $\bar{\mu}^{\prime }$ scheme}

\label{SEC4}

The construction of a phase space path integral in $\bar{\mu}^{\prime }$ scheme is more complex and subtler than in $\bar{\mu}$ scheme. As we illustrated in \ref{SEC2.2}, the new orthonormal basis in ${\cal H}_{Kin}^{grav}$ can be $\left| \lambda _1,\lambda _2,\lambda _3\right\rangle $ or $\left| \lambda _i,\lambda _j,v\right\rangle $. In order to compare with the $\bar{\mu}$ scheme situation conveniently, we choose the former one. Following the procedure employed in \ref{SEC3}, we are going to calculate the extraction amplitude $A\left( \vec{\lambda}_f,\phi _f;\vec{\lambda}_i,\phi _i\right) $.

The matter part of the extraction amplitude we use here is the same as in $\bar{\mu}$ scheme, so we ignore it for a little while and directly use it at
the end. Decompose the fictitious evolution into $N$ evolutions of length $\epsilon =\frac 1N$, the $n$-th term of the gravitational part is
\begin{eqnarray}
\left\langle \vec{\lambda}^{n+1}\right. \left| e^{\frac i\hbar \epsilon
\alpha \hat{C}^{grav}}\right| \left. \vec{\lambda}^n\right\rangle  &=&\delta
(\lambda _1^{n+1}-\lambda _1^n)\delta (\lambda _2^{n+1}-\lambda _2^n)\delta
(\lambda _3^{n+1}-\lambda _3^n)  \nonumber \\
&&+\frac i\hbar \epsilon \alpha \langle \lambda _1^{n+1},\lambda
_2^{n+1},\lambda _3^{n+1}|\hat{C}^{grav}|\lambda _1^n,\lambda _2^n,\lambda
_3^n\rangle +{\cal O}(\epsilon ^2).
\label{expC2}
\end{eqnarray}
As we known, although the number of possible values of $\lambda _i$ is countable, there does not exist an obvious superselection with respect to
it, therefore, we have to work with the entire positive real line it spans. So, for the possibility to perform a path integral, in a sense, we have used an approximation here. The Dirac delta is used rather than Kronecker delta to express the inner product $\left\langle \lambda _i^{n+1}\right. \left|
\lambda _i^n\right\rangle $, which means we will take $\lambda _i$ as continuous variables below. This assumption may generate a little confusion. When we think $\lambda _i$ is continuous, the quantum theory will be changed which caused by the changing of the Hilbert space. But, what we
interest in the following is the effective action, which is a semiclassical quantity itself. And consider the properties of $\lambda _i$, it is reasonable to perform a semiclassical approximation that let $\lambda _i$ to be continuous in the following calculations and do not affect the quantum theory. It is worth to emphasize that we will focus on the semiclassical situation from now on.

Then, make a algebraic transformation: $v^{\prime }=v+2$, $\lambda
_i^{\prime }=v^{\prime }\lambda _i/(v^{\prime }-2)$; $v^{\prime \prime }=v-2$%
, $\lambda _i^{\prime \prime }=v^{\prime \prime }\lambda _i/(v^{\prime
\prime }+2)$. In term of Eqs. (\ref{mubar'C}), (\ref{mubar'C1}) and (\ref{mubar'C2}%
), the matrix element of $\hat{C}^{grav}$ is expressed as
\begin{equation}
\left\langle \lambda _1^{n+1},\lambda _2^{n+1},\lambda _3^{n+1}\right.
\left| \hat{C}^{grav}\right| \left. \lambda _1^n,\lambda _2^n,\lambda
_3^n\right\rangle =\frac{\pi G}2\sqrt{v^nv^{n+1}}(v^{\prime }\langle
A\rangle +v^{\prime \prime }\langle B\rangle ),
\end{equation}
where
\begin{eqnarray}
\langle A\rangle  &=&\pm \left\langle \lambda _1^{n+1},\lambda
_2^{n+1},\lambda _3^{n+1}\right. \left| \lambda _1^n\pm \frac{\ell _0\hbar }{%
\lambda _2^{\prime n}\lambda _3^n},\lambda _2^{\prime n},\lambda
_3^n\right\rangle \pm \left\langle \lambda _1^{n+1},\lambda _2^{n+1},\lambda
_3^{n+1}\right. \left| \lambda _1^n\pm \frac{\ell _0\hbar }{\lambda
_2^n\lambda _3^{\prime n}},\lambda _2^n,\lambda _3^{\prime n}\right\rangle
\nonumber \\
&&\pm \left\langle \lambda _1^{n+1},\lambda _2^{n+1},\lambda _3^{n+1}\right.
\left| \lambda _1^n,\lambda _2^n\pm \frac{\ell _0\hbar }{\lambda _1^n\lambda
_3^{\prime n}},\lambda _3^{\prime n}\right\rangle \pm \left\langle \lambda
_1^{n+1},\lambda _2^{n+1},\lambda _3^{n+1}\right. \left| \lambda _1^{\prime
n},\lambda _2^n\pm \frac{\ell _0\hbar }{\lambda _1^{\prime n}\lambda _3^n}%
,\lambda _3^n\right\rangle   \nonumber \\
&&\pm \left\langle \lambda _1^{n+1},\lambda _2^{n+1},\lambda _3^{n+1}\right.
\left| \lambda _1^{\prime n},\lambda _2^n,\lambda _3^n\pm \frac{\ell _0\hbar
}{\lambda _1^{\prime n}\lambda _2^n}\right\rangle \pm \left\langle \lambda
_1^{n+1},\lambda _2^{n+1},\lambda _3^{n+1}\right. \left| \lambda
_1^n,\lambda _2^{\prime n},\lambda _3^n\pm \frac{\ell _0\hbar }{\lambda
_1^n\lambda _2^{\prime n}}\right\rangle ,
\end{eqnarray}
and
\begin{eqnarray}
\langle B\rangle  &=&\pm \left\langle \lambda _1^{n+1},\lambda
_2^{n+1},\lambda _3^{n+1}\right. \left| \lambda _1^n\mp \frac{\ell _0\hbar }{%
\lambda _2^{\prime \prime n}\lambda _3^n},\lambda _2^{\prime \prime
n},\lambda _3^n\right\rangle \pm \left\langle \lambda
_1^{n+1},\lambda _2^{n+1},\lambda _3^{n+1}\right. \left| \lambda _1^n\mp
\frac{\ell _0\hbar }{\lambda _2^n\lambda _3^{\prime \prime n}},\lambda
_2^n,\lambda _3^{\prime \prime n}\right\rangle   \nonumber \\
&&\pm \left\langle \lambda _1^{n+1},\lambda _2^{n+1},\lambda _3^{n+1}\right.
\left| \lambda _1^n,\lambda _2^n\mp \frac{\ell _0\hbar }{\lambda _1^n\lambda
_3^{\prime \prime n}},\lambda _3^{\prime \prime n}\right\rangle \pm
\left\langle \lambda _1^{n+1},\lambda _2^{n+1},\lambda
_3^{n+1}\right. \left| \lambda _1^{\prime \prime n},\lambda _2^n\mp \frac{%
\ell _0\hbar }{\lambda _1^{\prime \prime n}\lambda _3^n},\lambda
_3^n\right\rangle   \nonumber \\
&&\pm \left\langle \lambda _1^{n+1},\lambda
_2^{n+1},\lambda _3^{n+1}\right. \left| \lambda _1^{\prime \prime n},\lambda
_2^n,\lambda _3^n\mp \frac{\ell _0\hbar }{\lambda _1^{\prime \prime
n}\lambda _2^n}\right\rangle \pm \left\langle \lambda
_1^{n+1},\lambda _2^{n+1},\lambda _3^{n+1}\right. \left| \lambda
_1^n,\lambda _2^{\prime \prime n},\lambda _3^n\mp \frac{\ell _0\hbar }{%
\lambda _1^n\lambda _2^{\prime \prime n}}\right\rangle .
\end{eqnarray}
Employ the identity
\begin{equation}
\delta \left( \lambda ^{\prime }-\lambda \right) =\frac 1\pi \int_{-\infty
}^{+\infty }dke^{-ik\left(\lambda^{\prime}-\lambda\right)/2\hbar },
\end{equation}
where $k$ is the variable conjugated to $\lambda $ as we stated in Eq. (\ref{k}), we are led to the following expression for Eq. (\ref{expC2})
\begin{eqnarray}
&&\left\langle \vec{\lambda}^{n+1}\right. \left| e^{\frac i\hbar \epsilon
\alpha \hat{C}^{grav}}\right| \left. \vec{\lambda}^n\right\rangle   \nonumber
\\
&=&\frac 1{(\pi )^3}\int d\vec{k}^ne^{-i\vec{k}^n\cdot \left( \vec{\lambda}%
^{n+1}-\vec{\lambda}^n\right) /2\hbar }\left\{ 1+\frac i{2\hbar }\epsilon
\alpha \pi G\sqrt{v^nv^{n+1}}\right.   \nonumber \\
&&\times \left[ \left( v^{\prime }\right) ^n\left( e^{i\left(
-k_1^n\lambda _1^n\frac 1{v^{\prime n}}-k_2^n\lambda _2^n\frac 1{v^n}\right)
\ell _0}-e^{i\left( k_1^n\lambda _1^n\frac 1{v^{\prime n}}-k_2^n\lambda _2^n%
\frac 1{v^n}\right) \ell _0}\cdots +cyclic\,\,terms\right) \right.
\nonumber \\
&&\left. \left. +\left( v^{\prime \prime }\right) ^m\left( e^{i\left(
k_1^n\lambda _1^n\frac 1{v^{\prime \prime n}}+k_2^n\lambda _2^n\frac 1{v^n}%
\right) \ell _0}-e^{i\left( -k_1^n\lambda _1^n\frac 1{v^{\prime \prime n}}%
+k_2^n\lambda _2^n\frac 1{v^n}\right) \ell _0}\cdots
+cyclic\,\,terms)\right) +{\cal O}(\epsilon ^2)\right. \right\}.
\end{eqnarray}
Let $\ell _0k_i^n\lambda _i^n/v^n=a_i$, $\ell _0k_i^n\lambda _i^n/\left(
v^{\prime }\right) ^n=a_i^{\prime }$, and $\ell _0k_i^n\lambda _i^n/\left(
v^{\prime \prime }\right) ^n=a_i^{\prime \prime }$ for short. The
exponential terms of this equation can be expressed as
\begin{eqnarray}
&&v^{\prime }\left\{ \left[ \cos (a_1^{\prime }+a_2)-i\sin (a_1^{\prime
}+a_2)-\cos (a_1^{\prime }-a_2)-i\sin (a_1^{\prime }-a_2)\right]
+cyclic\,\,terms\right\}   \nonumber \\
&&+v^{\prime \prime }\left\{ \left[ \cos (a_1^{\prime \prime }+a_2)+i\sin
(a_1^{\prime \prime }+a_2)-\cos (a_1^{\prime \prime }-a_2)+i\sin
(a_1^{\prime \prime }-a_2)\right] +cyclic\,\,terms\right\} .
\label{sin}
\end{eqnarray}
Here, we introduce another approximation: assume the volume of the elementary cell is large enough, then we can let $v=v^{\prime }=v^{\prime\prime }$, and therefore the $i\sin $ terms in Eq. (\ref{sin}) are canceled with each other. So we have
\begin{eqnarray}
&&\left\langle \vec{\lambda}^{n+1}\right. \left| e^{\frac i\hbar \epsilon
\alpha \hat{C}^{grav}}\right| \left. \vec{\lambda}^n\right\rangle   \nonumber
\\
&=&\frac 1{(\pi )^3}\int d\vec{k}^ne^{-i\vec{k}^n\cdot \left( \vec{\lambda}%
^{n+1}-\vec{\lambda}^n\right) /2\hbar }\left\{ 1+\frac i\hbar \epsilon
\alpha \frac{\pi G}2\sqrt{v^nv^{n+1}}\left( -8v^n\right) \right.   \nonumber
\\
&&\left. \left. \times \left( \sin {\frac{k_1^n\lambda _1^n}{v^n}}\ell
_0\sin {\frac{k_2^n\lambda _2^n}{v^n}}\ell _0+\sin {\frac{k_2^n\lambda _2^n}{%
v^n}}\ell _0\sin {\frac{k_3^n\lambda _3^n}{v^n}}\ell _0+\sin {\frac{%
k_1^n\lambda _1^n}{v^n}}\ell _0\sin {\frac{k_3^n\lambda _3^n}{v^n}}\ell _0+%
{\cal O}(\epsilon ^2)\right) \right. \right\} .
\end{eqnarray}

Combining the scalar field part presented in Eq. (\ref{Cmatt}), the extraction
amplitude takes the form
\begin{eqnarray}
A(\vec{\lambda}_f,\phi _f;\vec{\lambda}_i,\phi _i) &=&\left\langle \vec{%
\lambda}^f,\phi ^f\right. \left| e^{\frac i\hbar \alpha \hat{C}%
^{grav}}\right| \left. \vec{\lambda}^i,\phi ^i\right\rangle   \nonumber \\
&=&\sum_{\vec{\lambda}^{N-1}\cdots \vec{\lambda}^1}\frac 1{(\pi )^{3N}}\int d%
\vec{k}_N\cdots d\vec{k}_1\cdot \left( \frac 1{2\pi }\right) ^{2N}\int
dp_N\cdots dp_1e^{\frac i\hbar S_N}+{\cal O}(\epsilon ^2),
\end{eqnarray}
where
\begin{eqnarray}
S_N &=&\epsilon \sum_{n=0}^{N-1}\left\{ p^{n+1}\frac{\phi ^{n+1}-\phi ^n}%
\epsilon -\frac{\vec{k}^n}2\cdot \frac{(\vec{\lambda}^{n+1}-\vec{\lambda}^n)}%
\epsilon +\alpha \left[ \left( p^n\right) ^2-4\pi G\sqrt{v^nv^{n+1}}%
v^n\right. \right.   \nonumber \\
&&\left. \left. \times \left( \sin {\frac{k_1^n\lambda _1^n}{v^n}}\ell
_0\sin {\frac{k_2^n\lambda _2^n}{v^n}}\ell _0+\sin {\frac{k_2^n\lambda _2^n}{%
v^n}}\ell _0\sin {\frac{k_3^n\lambda _3^n}{v^n}}\ell _0+\sin {\frac{%
k_1^n\lambda _1^n}{v^n}}\ell _0\sin {\frac{k_3^n\lambda _3^n}{v^n}}\ell
_0\right) \right] \right\} .
\end{eqnarray}
Finally, taking $N\rightarrow \infty $, we get
\begin{equation}
A\left( \vec{\lambda}^f,\phi ^f;\vec{\lambda}^i,\phi ^i\right) =\int d\alpha
\int \left[ D\vec{\lambda}(\tau )\right] \left[ D\vec{k}(\tau )\right]
\left[ Dp(\tau )\right] \left[ D\phi (\tau )\right] e^{\frac i\hbar S}
\label{F1}
\end{equation}
where
\begin{eqnarray}
S &=&\int_0^1d\tau \left\{ p\dot{\phi}-\frac 12\vec{k}\cdot \dot{\vec{\lambda}}%
-\alpha \left[ p^2-4\pi G\right. \right.   \nonumber  \label{F2} \\
&&\times v^2\left( \sin {\frac{k_1\lambda _1}v}\ell _0\sin {\frac{k_2\lambda
_2}v}\ell _0+\sin {\frac{k_2\lambda _2}v}\ell _0\sin {\frac{k_3\lambda _3}v}%
\ell _0\right.   \nonumber \\
&&\left. \left. \left. +\sin {\frac{k_1\lambda _1}v}\ell _0\sin {\frac{%
k_3\lambda _3}v}\ell _0\right) \right] \right\} .
\end{eqnarray}
Eq. (\ref{F1}) and Eq. (\ref{F2}) are the final expressions of the extraction amplitude and its action in $\bar{\mu}^{\prime }$ scheme. Notice that $\lambda _i$ has already been taken as a continuous variable, therefore, it is reasonable to interpret the $\left( \vec{\lambda}^{n+1}-\vec{\lambda}^n\right) /\epsilon $ as derivative here, and the integration ranges of $\vec{\lambda}$ and $\vec{k}$ are taken from $-\infty $ to $\infty $ directly. We no longer need to play the trick which was used in $\bar{\mu}$ scheme and the isotropic case to transform the integration range. Integration is over all trajectories in the classical phase space from the beginning.

\section{Discussion}

\label{SEC5} 

In this paper we extended the phase space path integral formulation of Friedmann space-times to anisotropic Bianchi I models, and performed the calculation both in $\bar{\mu}$ and $\bar{\mu}^{\prime }$ scheme. We restricted the matter source to be a massless scalar field and focused on the positive octant in which all three $p_i$ are positive.

Since $\bar{\mu}_i=\sqrt{\Delta/p_i}$ in $\bar{\mu}$ scheme, by setting $ \nu_i \sim p_i^{3/2}$, just as $\nu\sim p^{3/2}$ in the isotropic case, the
procedure used in \cite{PATH1} could be implemented directly. The formulation of the extraction amplitude then resembles three copies of the
isotropic model as we desired. To incorporate the $\bar{\mu}^{\prime }$ scheme, we have to overcome a few obstacles. The main problem comes from the fact
that the expression of the $\bar{\mu}^{\prime }_i$, which was given by Eq. (\ref{mubarp}), is much more complicated in $\bar{\mu}^{\prime }$ scheme than in $\bar{\mu}$ scheme.
This leads to the consequence that we have to make an algebraic simplification by introducing $\lambda_i\sim p_i^{1/2}$, and then the wave
function is dragged along the $\lambda_1$ direction by the unitary shift operator $E_1$. However, the affine distance involved in this dagging depends on $%
\lambda_2,\lambda_3$. From the difference equation of the gravitational part of the total constraint, it can be seen that the new variable $\lambda_i$
has a discrete spectrum but does not support on a specific $\lambda_i$-lattice. We can not write an integral representation for the Kronecker
delta by using a standard residue calculation like Eq. (\ref{Kroneckerdelta}).

In order to perform a path integral formulation in $\bar{\mu}^{\prime }_i$ case, we have used two approximations in \ref{SEC4}. The first one is taking $\lambda_i$
as continuous variable, then one can use the Dirac delta to express the inner product. The second one is to presume the volume of elementary cell is
large enough such that $(v-2)\approx v\approx (v+2)$. Actually they are the same thing. In general, our treatment presupposed that the the calculation
is away from the Planck regime, in other words, a semiclassical approximation was taken. This is the key point where $\bar{\mu}^{\prime }$ is different from $\bar{\mu}$ and the isotropic case. In the two later situations the calculation is exact and does not need any additional assumption.

If we compare the two path integrals in $\bar{\mu}$ and $\bar{\mu}^{\prime }$ scheme, it is obviously that the actions have the same formulation as in
isotropic case: a classical action plus a quantum gravity correction term, i.e., the $\sin $ term. Ignoring the constant terms, from Eq. (\ref{muS})
and Eq. (\ref{F2}), one can write down directly the two quantum gravity correction terms in the two schemes (denoted by $T$ and $T^{\prime }$ respectively). In $\bar{\mu}$ scheme,
\begin{eqnarray}
T &=&\nu ^1\nu ^2\sin \ell _0b^1\sin \ell _0b^2+\nu ^2\nu ^3\sin \ell
_0b^2\sin \ell _0b^3  \nonumber \\
&&+\nu ^1\nu ^3\sin \ell _0b^1\sin \ell _0b^3.
\end{eqnarray}
In $\bar{\mu}^{\prime }$ scheme,
\begin{eqnarray}
T^{\prime } &=&v^2\left( \sin {\frac{k_1\lambda _1}v}\ell _0\sin {\frac{%
k_2\lambda _2}v}\ell _0+\sin {\frac{k_2\lambda _2}v}\ell _0\sin {\frac{%
k_3\lambda _3}v}\ell _0\right.   \nonumber \\
&&\left. +\sin {\frac{k_1\lambda _1}v}\ell _0\sin {\frac{k_3\lambda _3}v}%
\ell _0\right) .
\end{eqnarray}
These terms do not look the same. By using the explicit expressions of $b_i,\lambda _i$ and $k_i$, $T$ and $T^{\prime }$ could be expressed as
\begin{equation}
T=\nu ^1\nu ^2\sin \bar{\mu}_1c_1\sin \bar{\mu}_2c_2+\nu ^2\nu ^3\sin \bar{%
\mu}_2c_2\sin \bar{\mu}_3c_3+\nu ^1\nu ^3\sin \bar{\mu}_1c_1\sin \bar{\mu}%
_3c_3,
\end{equation}
and
\begin{equation}
T^{\prime }=v^2\left( \sin \bar{\mu}_1^{\prime }c_1\sin \bar{\mu}_2^{\prime
}c_2+\sin \bar{\mu}_2^{\prime }c_2\sin \bar{\mu}_3^{\prime }c_3+\sin \bar{\mu%
}_1^{\prime }c_1\sin \bar{\mu}_3^{\prime }c_3\right).
\end{equation}
Notice that $\nu _i\sim p_i^{3/2}$ and $v\sim (p_1p_2p_3)^{1/2}$. Now we find that $T$ and $T^{\prime }$ are extremely similar in the formulation.
The difference is the immediate cause for the fiducial cell dependence on the semiclassical limit in $\bar{%
\mu}$ scheme. Consider the fiducial cell ${\cal V}$. Its volume is $V_0=L_1L_2L_3$. Notice that $c_i\sim L_i\dot{a}_i$ and $p_i\sim L_jL_ka_ja_k$, so if we rescale ${\cal V}$ as
\begin{equation}
V_0=L_1L_2L_3\rightarrow V_0^{\prime }=l_1L_1l_2L_2l_3L_3=l_1l_2l_3V_0,
\end{equation}
then
\begin{equation}
\bar{\mu}_1c_1\rightarrow \frac{l_1}{\sqrt{l_2l_3}}\bar{\mu}_1c_1,
\end{equation}
and
\begin{equation}
\bar{\mu}_1^{\prime }c_1\rightarrow \bar{\mu}_1^{\prime }c_1.
\end{equation}
This is the reason why $\bar{\mu}^{\prime }$ scheme has better scaling properties. These properties are already known in canonical quantization program. In path
integral formulation, we have proved again that the quantum dynamic does depend on the choice of the fiducial cell in $\bar{\mu}$ scheme, while it does not in
$\bar{\mu}^{\prime }$ scheme. Furthermore, if we take $p_1=p_2=p_3$, and therefore $\nu _i=v$, then we find $T=T^{\prime }\sim v^2\sin ^2\bar{\mu}c$. This is
precisely the quantum correction term in Friedmann model. Hence, it is rational to forecast that in the path integral formulation of Bianchi I
models the quantum bounce will also replace the big bang singularity due to the quantum gravity correction term. All of these properties provide a
strong evidence for the equivalence of the canonical approach and the path integral approach in LQC.

\acknowledgments This work was supported by the National Natural Science Foundation of China (Grant Nos. 11175019 and 11235003) and the Fundamental Research Funds for the Central Universities.


\begin{thebibliography}{99}
\bibitem{LQG} C. Rovelli, \emph{Quantum Gravity}, Cambridge University
Press, Cambridge (UK), 2004.

\bibitem{SF1} J. C. Baez, \emph{Spin foam models}, Class. Quantum Grav.
\textbf{15}, 1827(1998).

\bibitem{SF2} J. C. Baez, \emph{An introduction to spin foam models of BF
theory and quantum gravity}, Lect. Notes Phys. \textbf{543}, 25(2000).

\bibitem{LQC1} M. Bojowald, \emph{Loop quantum cosmology}, Liv. Rev. Rel.
\textbf{8}, 11(2005).

\bibitem{LQC-SF1} A. Ashtekar, M. Campiglia and A. Henderson, \emph{Loop
quantum cosmology and spin foams}, Phys. Lett. \textbf{B681}, 347-352(2009).

\bibitem{LQC-SF2} A. Ashtekar, M. Campiglia and A. Henderson, \emph{Casting
loop quantum cosmology in the spin foam paradigm}, Class. Quant Grav.
\textbf{27}, 135020(2010).

\bibitem{LQC-SF3} M. Campiglia, A. Henderson and W. Nelson, \emph{Vertex
expansion for the Bianchi I model}, Phys. Rev. \textbf{D82}, 064036(2010).

\bibitem{LQC-SF4} A. Henderson, C. Rovelli, F. Vidotto and E. Wilson-Ewing,
\emph{Local spinfoam expansion in loop quantum cosmology},
arXiv:1010.0502[gr-qc].

\bibitem{PATH1} A. Ashtekar, M. Campiglia and A. Henderson, \emph{Path
integrals and the WKB approximation in loop quantum cosmology}, Phys. Rev.
\textbf{D82}, 124043(2010) .

\bibitem{PATH2} Haiyun Huan, Yongge Ma and Li Qin, \emph{Path Integral and
Effective Hamiltonian in Loop Quantum Cosmology}, arXiv:1102.4755[gr-qc].

\bibitem{PATH3} Li Qin, Guo Deng and Yongge Ma, \emph{Path Integrals and
Alternative Effective Dynamics in Loop Quantum Cosmology}, Commun. Theor.
Phys. \textbf{57}, 326-332(2012)

\bibitem{BI1} D. W. Chiou, \emph{Loop quantum cosmology in Bianchi type I
models: Analytical investigarion}, Phys. Rev. \textbf{D75}, 024029(2007).

\bibitem{BI2} D. W. Chiou, \emph{Effective dynamics, big bounces and scaling
symmetry in Bianchi type I Loop Quantum Cosmology}, Phys. Rev. \textbf{D76},
124037(2007).

\bibitem{BI3} A. Ashtekar, E. Wilson-Ewing, \emph{Loop quantum cosmology of
Bianchi I models}, Phys. Rev. \textbf{D79}, 083535(2009).

\bibitem{AG} A. Ashtekar, J. Lewwandowski, \emph{Quantum theory of geometry:
I. Area operators}, Class. Quant Grav. \textbf{14}, A55-A81(1997).

\bibitem{BI4} {\L }. Szulc, \emph{Loop quantum cosmology of diagonal Bianchi
type I model: simplifications and scaling problems} Phys. Rev. \textbf{D78},
064035(2008).

\bibitem{BI6} M. Mart\'{\i}n-Benito, G. A. Mena Marug\'{a}n and T. Pawlowski,
\emph{Loop quantization of vacuum Bianchi I cosmology}, Phys. Rev. \textbf{D78},
064008(2008)

\bibitem{BI5} L. J. Garay, M. Mart\'{\i}n-Benito and G. A. Mena Marug\'{a}n, 
\emph{Inhomogeneous loop quantum cosmology: Hybrid quantization of the Gowdy model}, 
Phys. Rev. \textbf{D82}, 044048(2010)

\end{thebibliography}
\end{document}